\documentclass[12pt,emtex]{article}
\usepackage{amsfonts}
\usepackage{amssymb}
\usepackage{amscd}
\usepackage{amsmath}
\newcommand{\BEQ}{\begin{equation}}
\newcommand{\EEQ}{\end{equation}}

\newtheorem{Th}{Theorem}

\newtheorem{Lem}{Lemma}

\newtheorem{Cor}{Corollary}

\newtheorem{Not}{Notation}
\newtheorem{Rem}{Remark}

\def\nn{\nonumber}

\def\bea{\begin{eqnarray}}
\def\eea{\end{eqnarray}}

\def\CC{{\mathbb{ C}}}
\def\RR{{\mathbb{ R}}}

\def\g{\mathfrak{gl}}
\def\s{\mathsf{symb}}

\def\K{\mathcal{K}}

\def\one#1{#1^{\raise5pt\hbox{$\scriptstyle\!\!\!\!1$}}\,{}}
\def\two#1{#1^{\raise5pt\hbox{$\scriptstyle\!\!\!\!2$}}\,{}}

\begin{document}

\begin{titlepage}
\hfill ITEP-TH-105/03 \vskip 2.5cm

\centerline{\LARGE \bf Rational Lax operators and their quantization}

\vskip 1.0cm
\centerline{A. Chervov \footnote{E-mail:
chervov@itep.ru} }
\centerline{\sf Institute for Theoretical and Experimental Physics
}

\vskip 1.0cm

\centerline{ L. Rybnikov \footnote{E-mail:
leo\_rybnikov@mtu-net.ru
} }

\centerline{\sf Moscow State University}

\vskip 1.0cm

\centerline{ D. Talalaev \footnote{E-mail:
talalaev@gate.itep.ru}
\footnote{The work of A.C. and D.T.,
has been partially supported by the RFBR grant 01-01-00546,
the work of A.C. was partially supported by the
Russian President's grant 00-15-99296, the work of D.T.
has been partially supported by the RFBR grant for the support
of young scientists MAC-2003 N 03-01-06236 under the
project 01-01-00546.
}
}

\centerline{\sf Institute for Theoretical and Experimental Physics
}

\vskip 2.0cm

\centerline{\large \bf Abstract}

\vskip 1.0cm

We investigate the construction of the quantum commuting hamiltonians
for the Gaudin integrable model.
We prove that $[Tr L^k(z), Tr L^m(u) ]=0$, for $k,m < 4 .$ However this
naive receipt of quantization of classically commuting hamiltonians
fails in general, for example we prove that
$[Tr L^4(z), Tr L^2(u) ] \ne 0$.
We investigate in details the case of the one spin Gaudin model
with the magnetic field also known as the model obtained by the
"argument shift method".
Mathematically speaking this method gives maximal Poisson commutative
subalgebras in the symmetric algebra $S(\g(N)).$
We show that such subalgebras can be lifted to $U(\g(N)),$ simply
considering $Tr L(z)^k$, $k\le N$ for $N<5$. For $N=6$ this method fails:
$[Tr L_{MF}(z)^6, L_{MF}(u)^3]\ne 0 $.
All the proofs are based on the explicit calculations using
$r$-matrix technique.
We also propose the general receipt
to find the commutation formula for powers of Lax operator.
For small power exponents we find the complete
commutation relations between powers of Lax operators.


\end{titlepage}

\tableofcontents

\section{Introduction}

This paper is devoted to the investigation of rational Lax operators of
Gaudin type on quantum the level. The Gaudin model was introduced by Gaudin in
\cite{Gaudin} (see section 13.2.2) as a limit of the famous XXX-Heisenberg
model, which describes interaction of spins in one dimensional chain.
It appears to be that Gaudin model related to various
fields of research in mathematics and mathematical physics:
Knizhnik-Zamolodchikov equation \cite{KZ} and isomonodromy deformation
theory (see \cite{MAO-AML});
Hitchin  system \cite{Hit} (see  \cite{NN}, also \cite{ER1,GAW});
Langlands correspondence (see \cite{Frenkel});
geometry of polygons (see \cite{Milson,Falqui}).
Also it seems to be
not so widely known that integrable systems obtained by
argument shift method \cite{Mishen-Fomen}, are
the most simplest particular (one spin) cases of Gaudin model, more precisely
one should add to the Gaudin Lax operator constant matrix,
(physically this means turning on the magnetic field).

But despite lots of results concerning the Gaudin model
 it seems that some simple questions remain still
open. One of such questions is explicit construction of higher Gaudin's
hamiltonians on the quantum level. On the classical  level the Gaudin
hamiltonians can be obtained as values at different $z$
of $Tr L^k(z)$, where $L(z)$ is the Lax operator (see below). They are not
independent but the it can be chosen a basis in this family of
functions composed by coefficients of expansions at poles. We
investigate the question whether the same construction on the quantum
level gives commuting hamiltonians, we prove that
$ [Tr L^k(z), Tr L^m(u)]= 0 $ for $k\le 3, m \le 3, \forall z,u$, but
$ [Tr L^4(z), Tr L^2(u)] \ne 0 $. So by considerations
of $Tr L^k(z)$ one can find enough commuting hamiltonians
for $\g(2), \g(3) $, but not for the higher rank.
We also find some general formulas for the
commutation between  matrix elements of powers $(L^k(z))_{i,j}$.
The special case of the one spin Gaudin model
(equivalent to the model constructed by the argument shift method)
is more optimistic, we obtain:
$ [Tr L^k(z), Tr L^2(u)]= 0 $ for $ \forall k,z,u$,
$ [Tr L^k(z), Tr L^3(u)]= 0 $ for $ k\le 5 \forall z,u$,
but $ [Tr L^6(z), Tr L^3(u)] \ne 0 $, which means that in
this particular case one is also unable to quantize
the Poisson commuting subalgebra by considering the
$Tr L^k(z) $, starting from $\g(6)$.

Let us mention that the problem of lift
of Poisson commutative subalgebra generated
by the argument shift method has been
investigated before:
in \cite{NazO}
the method of quantization was proposed on
the basis of Yangian technique,
another receipt was proposed in \cite{Skr},
and it's also worth to mention that in \cite{Tarasov} the Vinberg's
conjecture (see \cite{Vinberg}) that
the  symmetrization map from $S(\g(n))\to U(\g(n))$
provides the necessary quantization was proved.
Despite all these results it seems that the
solution of the problem is not in the
stage as one can hope:
one can hope for some simple formula
like $Tr L^k(z)$ or some modification of it
for the  quantization of such subalgebras.

The same can be said about the Gaudin model:
in \cite{Frenkel,Feigin-Frenkel,Feigin-Frenkel-Resh}
it was shown that the Gaudin hamiltonians
can be obtained from the center
on the critical level
of the
universal enveloping of the corresponding
Kac-Moody algebra.
It was proved  in \cite{Feigin-Frenkel}
that the center on the critical level
is big enough, but there is no explicit
construction of the center.
On the other hand one can try to obtain
the higher Gaudin hamiltonians from the
known commuting hamiltonians
of the XXX-Heisenberg spin chain.

Let us present the result of our paper in more explicit way. Let us recall
some notations first: let us denote by $\Phi$ the following $n\times
n$-matrix with coefficients in $\g(n)$: we put element $e_{ij}$ on
$ij$-th place of the matrix. We also consider the direct sum $\g(n)\oplus
... \oplus \g(n)$ and denote by $\Phi_i$ the matrix defined as above but
with the elements from the  $i$-th copy of $\g(n)$ in $\g(n)\oplus ...
\oplus \g(n)$. Let us introduce the Lax operator for the Gaudin model:
\bea \label{i-Lax-Gaud} L(z)=\mathcal{K}+\sum_{i=1...N}
\frac{\Phi_i}{z-a_i} \eea where $a_i\in \CC, \Phi_i ....$,  $ \mathcal{K}$
arbitrary constant matrix, (physically  $ \mathcal{K}$  corresponds to a
magnetic field.)

\bea
\label{i-tr-l2}
Tr L^2(z) =
\sum_i \frac{Tr \Phi_i^2}{(z-a_i)^2}
+\sum_k \frac{1}{z-a_k}
(Tr \K \Phi_k +\sum_{j\ne k} \frac{2 Tr \Phi_k \Phi_j}{(a_k-a_j)})
\eea

\bea
\label{i-ham-gaud}
H_k=Tr \K \Phi_k + \sum_{j\ne k} \frac{2 Tr \Phi_k \Phi_j}{(a_k-a_j)}
\in U(\g(n))\otimes ... \otimes U(\g(n)) \\
\mbox{ - are called quadratic Gaudin Hamiltonians }\nn
\eea

It's known that $ [Tr L^2(u), Tr L^2(z)]=0 $ and so that $[H_k, H_j]=0$.
The Gaudin model on quantum level consists of taking some representation
$V_{1}\otimes...\otimes V_{N}$ of
$\g(n)\oplus ... \oplus \g(n)$ and asking
for the spectrum of such operators, their matrix elements etc.

On the classical level one uses the symbol  map $\s:U(\g(n))\to S(\g(n))$
and so the images of $H_i$ are functions on the $\g(n)^*\oplus ... \oplus
\g(n)^*$ so one can restrict them to the orbits ${\cal O}_1\times...\times
{\cal O}_N $, and so one obtains the phase space ${\cal
O}_1\times...\times {\cal O}_N$ with Kirillov's symplectic form and
functions $\s(H_i)$ each of which can be taken as hamiltonian, so the
classical hamiltonian systems is described. It's known that it is
completely integrable: one can take coefficients at $(z-a_i)^{-m}$
of  $\s(Tr L^k(z))$ as hamiltonians which will Poisson commute and (at
least for the general choice of orbits, it seems that the question of
considering different non-general orbits has not been analyzed carefully)
this will give the number of independent hamiltonians equal to the half of
the dimension of the phase space which means the complete integrability in
the Liouville sense.

The fact that $Tr (L^k(z))$ and $Tr (L^l(u))$ commute  with each
other for any $z,u,k,l$ with respect to the Poisson bracket can be
proved by the r-matrix technique on the basis of commutation relation
between $(L^k(z))_{kl}, (L^l(u))_{qp}$.
Leningrad's notation seems to be very useful for such calculation: we denote by
$\one T= T\otimes Id, \two T= Id \otimes T$.
So $[\one A , \two B]_{ij,kl}=[a_{ij},b_{kl}] $,
hence the commutation relation
between $\one A$ and $ \two B$ encodes all the
commutation relations between
all $a_{ij}$  and $b_{kl}$.
Matrix $P$ is defined by the rule: $P(a\otimes b )=b \otimes a $.

Our first results are the following formulas for the commutation relation
between powers of Lax operator $L(z):$
\bea
\label{i-ln-l}
 [ \one L^n(z), \two L(u) ] &=&
\two L^n(z) \frac{P}{z-u}  - \one L^n(z)  \frac{P}{z-u}+ \nn \\ ~~
&&\sum_{i=0}^{n-1} \one L^i(z) (\one L(u) - \two L(u)) \two L^{n-1-i}(z)
\frac{P}{z-u}
\eea
\bea
\label{i-ln-l2}
[ \one L^n(z), \two L^2(u) ] &=&
(\one L(u)  + \two L(u) ) \two L^n(z) \frac{P}{z-u}
-\one L^n(z) (\one L(u) + \two L(u))  \frac{P}{z-u}+ \nn \\ ~~
&&+\sum_{i=0}^{n-1}
\one L^i(z) ( \one L^2(u)   - \two L^2(u) ) \two L^{n-1-i}(z) \frac{P}{z-u}
\eea
\bea
 \label{i-ln-l3}
[ \one L^n(z), \two L^3(u) ] &=&
\sum_{i=0}^{n-1}
\one L^i(z) ( \one L^3(u)   - \two L^3(u) ) \two L^{n-1-i}(z) \frac{P}{z-u}
+ \nn \\ ~~
&& + (\one L^2(u)+ \one L(u) \two L(u)  + \two L^2(u) ) \two L^n(z)
\frac{P}{z-u} \nn \\ ~~
&&
-\one L^n(z) (\one L^2(u)+ \one L(u) \two L(u)  + \two L^2(u) )  \frac{P}{z-u}+ \nn \\ ~~
&& + [\frac{\partial}{\partial u} \one L(u) +\frac{\one L(u)}{z-u}, \one
L^n(z)] \frac{1}{z-u}
\eea
\bea
\label{i-ln-lm}
[ \one L^n(z), \two L^m(u) ] &=&
\sum_{i=0}^{n-1}
\one L^i(z) ( \one L^m(u)   - \two L^m(u) ) \two L^{n-1-i}(z) \frac{P}{z-u}
+ \nn \\ ~~
&&+( \sum_{i=0}^{m-1} \one L^i(u)\two L^{m-1-i}(u) ) \two L^n(z) \frac{P}{z-u}
- \nn \\ ~~
&&- \one L^n(z) ( \sum_{i=0}^{m-1} \one L^i(u)\two L^{m-1-i}(u) )
\frac{P}{z-u} + \nn\\
&& + \mbox{  terms of lower degree }
\eea

{\Rem Let us emphasize the importance of such formulas. The special kind of
normal ordering adapted to calculation of traces is realized therein.
Indeed, in the right hand side we see the sums of terms
of the type $\one A_1 \one A_2 ... \one A_n \two B_1 \two B_2 ... \two B_l
P$, and  there are no terms of the type: $Tr \one A \two B \one C \two D
\one E P $. We need such formulas because our main aim is to investigate
the commutators of the form $ [Tr L^k(z)  , Tr L^p(u) ]$ which can be
represented by the formula: $ [Tr A, Tr B ] = Tr [ \one A, \two B] $. And
it is easy to see that $ Tr \one A_1 \one A_2 ... \one A_n \two B_1 \two
B_2 ... \two B_l P= Tr (A_1A_2...A_nB_1B_2...B_l)$, but there is no simple
formula for the expression like $Tr \one A \two B \one C \two D \one E $.}

On the basis of the formulas above we are able
to obtain the following results on the commutativity of traces:
\bea
\label{i-tr-tr}
[ Tr L^k(z), Tr L^l(u)] & = & 0 \mbox{ for $k\le 3$, $l\le 3$} \forall z,u \\
\label{i-tr4-tr2}
[ Tr L^4(z), Tr L^2(u)] & \ne & 0 \\
\label{i-tr}
 Tr [ L^k(z),  L^l(u)] & = & 0 \mbox{ for $k\le 3$, $l\le 2$} \forall z,u \\
\label{i-tr4-2}
Tr [ L^4(z),  L^2(u)] & \ne & 0 \\
\eea
(Let us mention that it can be seen from the proofs that
$[ Tr L^k(z), Tr L^l(u)]  =  0$ is more or less equivalent to the
$ Tr  [ L^k(z),  L^{l-1} (u)]  =  0 $.)

We pay special attention to the case of the one pole Gaudin model (the system
of Mishchenko-Fomenko):

\bea \label{i-Lax-mf} L_{MF}(z)=\mathcal{K}+ \frac{\Phi}{z} \eea
Considerations of $\s[Tr L^k(z)]$ gives Poisson commutative subalgebra
in $S(\g(n))$. This method is known as the ``argument shift method''
\cite{Mishen-Fomen}. We analyze to what extent $Tr L^k(z)$ can be used to
obtain commutative subalgebras in $U(\g(n))$. We obtain that even in this
simplest case there is no commutativity on the quantum level at least for
$\g(n)$, $n\ge6$: \bea [ Tr  L_{MF}^k(z), Tr L_{MF}^2(z)]&=& 0 ~~ \forall
k  \\ ~~ [ Tr  L_{MF}^k(z), Tr L_{MF}^3(z)]&=& 0 ~~ for ~~ k \le 5  \\ ~~
[ Tr  L_{MF}^6(z), Tr L_{MF}^3(z)]& \ne& 0 \eea

We also prove that subalgebra generated by $Tr L^k(z)$
contains the Cartan subalgebra, moreover it commutes with
the Cartan subalgebra.

{\bf Acknowledgements.} We are indebted to A. Molev, E. Sklyanin,
A. Zotov  for useful discussions and correspondence;
to B. Jurco  for sending us his paper \cite{Jurco}.
Let us also acknowledge reading very useful surveys
\cite{Molev,Skl1,Skl2}.

\section{Preliminaries}

\subsection{Leningrad's notation}

Let us denote by $P$
 the transposition matrix in the tensor product, so it
acts as follows $$P (v_1\otimes v_2)=v_2\otimes v_1$$ and in terms
of matrix elements it could be expressed as:
$$P=\sum_{i,j}e_{ij}\otimes e_{ji}.$$

{\Not Leningrad's notation $\one T=T\otimes Id , \two T = Id \otimes T$.
}

Always it is meant the following index notation for an element of
$GL_n\otimes GL_n\otimes \mathcal{A}$ where $\mathcal{A}$ is an associative
algebra
$$C=\sum_{ij,kl}C_{ij,kl}e_{ij}\otimes e_{kl}$$ where $e_{ij}$ are
generators of  $GL_n.$ Then  $(\one T)_{ij,kl}=T_{ij}\delta_{kl}$ and
$(\two T)_{ij,kl}=\delta_{ij} T_{kl}.$

Let us mention the following  useful properties of such notation:
\bea \label{A1-B2} [\one A, \two B ]_{ij,kl} = [A_{ij}, B_{kl} ] \eea
hence one matrix commutation relation between $\one A$ and $ \two B$ encodes all
the commutation relations between all matrix elements $a_{ij}$  and $b_{kl}$.
We will see later (see formula \ref{fml-com1}) that the
commutation relations between basic elements of the Lie algebra $\g(n)$
are encoded in one simple relation, moreover this relation does not depend
on $n$.

Let us mention the following useful properties of such notation:

\bea
\label{P1}
&P \one A \two B = \two A \one B P& \\
\label{P2}
&P [\one A, P] = -[ \one A,  P ] P= [P, \one A ] P= [\two A, P]P = \one A - \two A &\\
&P[ \one A, P]=[P, \two A] P  & \\
\label{Tr1}
& Tr P  \one A \two B = Tr   \one A \two B P  =
Tr  P  \two A \one B   = Tr    \two A \one B P   = Tr (AB) & \\
\label{Tr2}
&[ Tr A, Tr B]= Tr [ \one A, \two B ]  \mbox {~~~~~~~ (Main useful property I) } &\\
\label{Tr3}
& Tr [ A, B]= Tr [ \one A, \two B ] P  \mbox{ ~~~~~~~ (Main useful property II) }  &
\eea

For any matrix $X_{ij,kl}$ in tensor square of some space $V$, the
following is true:
\bea
\label{XP}
 (X P)_{ij,kl}=X_{il,kj}\\
\label{PX}
 (PX)_{ij,kl}=X_{kj,il}
\eea

\subsection{Formulas without spectral parameter}

Let us recall the basic facts from the
r-matrix technique in the case of r-matrix without
spectral parameter. We do this
because our main aim will be to obtain analogous
results in the case or L-operator
of the Gaudin model
with spectral parameter.

Introduce matrix $\Phi$ which is matrix with coefficients
in $U(\g(n))$, defined by the following rule:
we just put the element $e_{ij}$ on $ij$-th place of the matrix.
It's the simplest prototype of L-operators in integrable
systems.

{\Lem \label{l1} \label{com1} The commutation relations between elements
$e_{ij}\in \g(n):$
$$[e_{ij},e_{kl}]=e_{il}\delta_{jk}-e_{kj}\delta_{li}$$ can be encoded in
the following way using Leningrad's notation: \bea \label{fml-com1}
  [\one \Phi ,  \two \Phi ]= [\one \Phi, P]=[P, \two \Phi]=\frac{1}{2}
[\one \Phi - \two \Phi, P]
\eea
}
\\
This commutation relation is of the so-called "r-matrix" type
without spectral parameter.
Matrix $P$ is the simplest example of the classical
r-matrix.

{\Cor \label{C1}
The formula above can be rewritten as
\begin{equation}\label{com1C}
  [\one \Phi - \frac{1}{2}P ,  \two \Phi - \frac{1}{2}P]=0
\end{equation}
this can be used to obtain formulas like
$  [(\one \Phi - \frac{1}{2}P)^l ,  (\two \Phi - \frac{1}{2}P)^n]=0$,
but we will not use this here.
}
\\
For any matrix $\Phi$ satisfying the formula in lemma \ref{l1}
the following is true:

{\Lem \label{l2} By the Leibnitz rule one immediately obtains:
\begin{equation}\label{com2}
  [\one \Phi^n ,  \two \Phi ]= [\one \Phi^n, P]=[P, \two \Phi^n]
\end{equation}
}

{\Lem \label{l3} Using the formula above one obtains:
\begin{equation}\label{com3}
  [\one \Phi^r ,  \two \Phi^s ]= \sum_{a=1}^{min(r,s)} P \one \Phi^{a-1}
\two \Phi^{r+s-a} - P \one \Phi^{r+s-a} \two \Phi^{a-1}
\end{equation}
}
\\
As a demonstration of the "r-matrix" technique
we prove the following statement, which is due to
Gelfand \cite{Gelfand}.

{\Lem The elements $Tr \Phi^r \in U(\g(n))$ lie in the center of the
$U(\g(n))$, i.e. they are Casimirs of $U(\g(n))$. }
\\
{\bf Proof~}
\bea
[ Tr \Phi^r, \Phi_{i,j} ]= \sum_{k}
([ \one \Phi^r, \two \Phi])_{kk,ij}=
\sum_{k} ([ \one \Phi^r, P ] )_{kk,ij}
=
\sum_{k}( \one \Phi^r P - P \one \Phi^r)_{kk,ij}=\nn\\
=
\sum_{k}( \one \Phi^r )_{kj,ik}  -  (\one \Phi^r)_{ik,kj}
=
\sum_{k} (\Phi^r_{kj} \delta_{ik} -\Phi^r_{ik} \delta_{kj})
=  ( \Phi^r_{i,j}  -\Phi^r_{i,j} )=0
\eea
$\Box$

{\Rem~} In fact $Tr \Phi ^r$, $r=1,...,n$
generate the center of $U(\g(n))$ (see \cite{Gelfand}).


{\Lem \label{l4} One can also prove the following
(see for example \cite{Molev}):
\begin{equation}\label{com4}
  [\one \Phi^{r+1} ,  \two \Phi^s ] -    [\one \Phi^{r} ,
\two \Phi^{s+1} ]
=  P \one \Phi^{r} \two \Phi^{s} - P \one \Phi^{s} \two \Phi^{r}
\end{equation}
}


\subsection{Gaudin model and its Lax operator with spectral parameter}


Let us recall the Gaudin model. Consider the direct sum
$\g(n)\oplus ... \oplus \g(n)$ and  denote by $\Phi_i$ the matrix $n\times n$
with values in this direct sum defined as follows:
$(\Phi_i)_{kl}$ is the $e_{kl}$ generator of the $i$-th copy of
$\g(n)$ in the direct sum $\g(n)\oplus ... \oplus \g(n).$
Let us introduce the
Lax operator for the  Gaudin model: \bea \label{Lax-Gaud}
L(z)=\mathcal{K}+\sum_{i=1...N} \frac{\Phi_i}{z-a_i} \eea where $a_i\in
\CC$,  $ \mathcal{K}$ is an arbitrary constant matrix (in original
physical applications  the case of $sl(2)$ was important, in this case
variables $\Phi$ corresponds to spins and  $ \mathcal{K}$  corresponds to
magnetic field. Nowadays the range of physical applications of integrable
spin chains is quite reach
and includes not only $sl(2)$ case, see for example \cite{Zarembo, Gorsky}).
\\
Let us consider
\bea
\label{tr-l2}
Tr L^2(z) =
\sum_i \frac{Tr \Phi_i^2}{(z-a_i)^2}
+\sum_k \frac{1}{z-a_k}
(Tr \K \Phi_k +\sum_{j\ne k} \frac{2 Tr \Phi_k \Phi_j}{(a_k-a_j)})
\eea

\bea
\label{ham-gaud}
H_k=Tr \K \Phi_k + \sum_{j\ne k} \frac{2 Tr \Phi_k \Phi_j}{(a_k-a_j)}
\in U(\g(n))\otimes ... \otimes U(\g(n)) \\
\mbox{ - are called quadratic Gaudin Hamiltonians }\nn
\eea
\\
It's known that $ [Tr L^2(u), Tr L^2(z)]=0 $ and so that $[H_k, H_j]=0$.
The Gaudin model on quantum level consists of taking some representation
$V_{1}\otimes...\otimes V_{N}$ of
$\g(n)\oplus ... \oplus \g(n)$ and asking
for the spectrum of such operators, their matrix elements etc.

On the classical level one uses the symbol  map
$\s:U(\g(n))\to S(\g(n)).$
The images of $H_i$ can be interpreted as functions on
$\g(n)^*\oplus ... \oplus \g^*(n).$ One restricts them to the
product of coadjoint orbits ${\cal O}_1\times...\times {\cal O}_N $ which
is by definition the phase space of classical Gaudin model. The
symplectic structure on this space is  Kirillov's symplectic form. For the
hamiltonian one can take any of functions $\s(H_i)$. The classical Gaudin model
is known to be completely integrable: one can take coefficients at
${(z-a_i)^{-m}}$ of  $\s(Tr L^k(z))$ as hamiltonians which
Poisson commute and at least for the general choice of orbits (though it
seems that the question of considering different non-general orbits has
not been analyzed carefully) this gives the family of independent
hamiltonians. Their number equals to the half the dimension of the phase space which
means the complete integrability in the Liouville sense.

More mathematically speaking one should mention
that Gaudin hamiltonians generate the
maximal Poisson-commutative subalgebra in
$S(\g(n))\otimes ... \otimes S(\g(n))$, though it's difficult to give
the reference for this result.

The Lax-pair representation for the
Gaudin model on the both quantum
and classical level
 can be found in \cite{Jurco}.
The Gaudin Lax operator was generalized
to the trigonometric \cite{Jurco} and elliptic
\cite{Tak-Skl} dependence on the spectral
parameter. It was also generalized
to include higher poles in $z$
see \cite{Beauv,AHH,Chern}.

\section{Commutativity and noncommutativity of traces }

\subsection{Commutation relations with spectral parameter}

As it was mentioned before it's known that Gaudin model is completely
integrable on the classical and quantum  levels, moreover the commuting
hamiltonians on the classical level can be given as coefficients
at $(z-a_i)^{-m}$ of $\s(Tr L(z)^k) $, but no explicit formula for
the higher than quadratic quantum integrals is known. Our aim is to check whether
the $Tr L(z)^k $ commute on the quantum level. Before doing this it is
necessary to obtain the commutation relations between powers of $L(z)$,
generalizing such relations in the case without spectral parameter
see formulas \ref{com2},\ref{com3}.

{\Lem \label{r-mat}
Using formula \ref{fml-com1}
one immediately obtains:
\bea
\label{fml-r-mat}
[ \one L(z), \two L(u) ] =[\one L(u), \two L(z)]=
[\frac{P}{z-u},  \one L(z) + \two L(u) ]= \nn \\
= [\frac{P}{z-u},  \one L(z) - \one L(u) ] = [\frac{P}{z-u},  -\two L(z) + \two L(u) ]
\eea
}
\\
This commutation relation is of the so-called linear r-matrix type. The
matrix $\frac{P}{z-u}$ is the simplest rational r-matrix. In most of
known integrable systems the Lax operator satisfies an analogous relation (with
the other r-matrices and possibly with the "linear" relation changed to
the "quadratic" one). These relations provide a simple construction
on the classical and sometimes on the quantum level.

{\Cor \label{C2}
The formula above can be rewritten as
\begin{equation}\label{r-matC}
[ \one L(z) - \frac{P}{z-u}, \two L(u) + \frac{P}{z-u} ] =0.
\end{equation}
}

{\Lem \label{r-mat-yang-bax}
The r-matrix $r(z,u)= \frac{P}{z-u} $
 satisfies classical Yang-Baxter equation:
\bea
[r_{12}(u_1,u_2), r_{13}(u_1,u_3)+r_{23}(u_2,u_3)]
-[r_{13}(u_1,u_3), r_{32}(u_3,u_2)]=0 \in V\otimes V\otimes V
\eea
where we use the standard notation $r_{ij}$ which means
linear operator in
$V\otimes V\otimes V$
 which act as operator $r$, but the
action is on i-th and j-th components of the tensor
product $V\otimes V\otimes V$,
for example $P_{13} (a\otimes b\otimes c)=  (c\otimes b\otimes a)$.
}

 {\Rem ~} There is some confusion in notation: one  says "classical
r-matrix", "classical Yang-Baxter equation" though we deal with the
quantum problem (commutators instead of Poisson brackets). This is due to
the historical reasons: in the case of quadratic commutation relations
(which seems to be appeared before linear relations) quantum R-matrix
corresponds to the quantum case and classical to the classical case. In
the case of linear commutation relations classical Yang-Baxter equation
and classical r-matrix corresponds to both classical and quantum cases.
\vskip 5mm
To go further we need for several technical lemmas:
{\Lem \label{com-rel-u-u}
For the Gaudin Lax it's true that:
\bea
[ \one L(u), \two L(u) ] =[P,\frac{\partial}{\partial u} \one L(u) ]
\eea
}

{\Lem \label{l-ln-easy} One has the following expressions for
commutators:
\bea
[ \one L(z), \two L(u) ] &=&
(\one L(u)  - \two L(u) ) \frac{P}{z-u}+
(\two L(z) - \one L(z) ) \frac{P}{z-u} \\ ~~
 [ \one L(z), \two L^2(u) ] & =&
(\one L^2(u)  - \two L^2(u) ) \frac{P}{z-u}+ \nn \\ ~~
&&\two L(u) (\two L(z)  - \one L(z) ) \frac{P}{z-u} +
(\two L(z) - \one L(z)) \one L(u) \frac{P}{z-u}  \\ ~~
 [ \one L(z), \two L^n(u) ] &=&
(\one L^n(u)  - \two L^n(u) ) \frac{P}{z-u}+ \nn \\ ~~
&&\sum_{i=0}^{n-1} \two L^i(u) (\two L(z) - \one L(z)) \one L^{n-1-i}(u)
\frac{P}{z-u}
\eea
}

{\Lem \label{ln-l-easy} The expressions of the previous lemma can
be rewritten in the ordered form\footnote{
Our aim is to obtain the formula
for $ [L^n(z), L^m(u)]$
in the form which is the sum of
terms like $$\one A_1 \one A_2 ... \one A_n
\two B_1 \two B_2 ... \two B_l P^{0~or~1}$$
and  there are no terms of the disordered type:
$Tr \one A \two B \one C \two D \one E P $.
We need such formulas because they are well adapted to calculation of
 $ [Tr L^k(z)  , Tr L^p(u) ],$ indeed,
$ [Tr A, Tr B ] = Tr [ \one A, \two B] $ and it is easy to see that
$$ Tr \one A_1 \one A_2 ... \one A_n
\two B_1 \two B_2 ... \two B_l P= Tr (A_1A_2...A_nB_1B_2...B_l)$$
but there is no simple formula for the expression like
$Tr \one A \two B \one C \two D \one E $.
}:
\bea
[ \one L(z), \two L(u) ] &=&
(\one L(u)  - \two L(u) ) \frac{P}{z-u}+
(\two L(z) - \one L(z) ) \frac{P}{z-u} \\ ~~
 [ \one L^2(z), \two L(u) ] & =&
(\two L^2(z)  - \one L^2(z) ) \frac{P}{z-u}+ \nn \\ ~~
&& \one L(z) (\one L(u)  - \two L(u) ) \frac{P}{z-u}+
(\one L(u) - \two L(u)) \two L(z) \frac{P}{z-u}
 \\ ~~
\label{fml-ln-l-easy}
 [ \one L^n(z), \two L(u) ] &=&
(\two L^n(z)  - \one L^n(z) ) \frac{P}{z-u}+ \nn \\ ~~
&&\sum_{i=0}^{n-1} \one L^i(z) (\one L(u) - \two L(u)) \two L^{n-1-i}(z)
\frac{P}{z-u}
\eea
}
\\
{\bf Receipt to obtain the formulas for
$ [L^n(z), L^m(u)]$.}
Let us formulate the main necessary observation which provides
such formulas. It is easy to obtain the
commutation relation
 between $\one L^n(z), \two L(u) $
in the form presented above: i.e. as the
sum terms of the form $\one A_1 ... \one A_p
\two B_1 ... \two B_q P $, but it is difficult
to obtain the same answer as the sum of terms
of the form $\two B_1 ... \two B_q \one A_1 ... \one A_p
 P $.  {\em As soon as one is able to
find such formula then the formula
for the  commutator of $\one L^n(z), \two L^m(u) $ follows immediately.}

The reason is quite simple:
assume that $[\one A, \two B ] = \one C \two D P$
then
$$[\one A^n, \two B ] = \sum_{i=0}^n \one A^i \one C \two D P \one A^{n-1-i}=
\sum_{i=0}^n \one A^i \one C \two D  \two A^{n-1-i}. P$$
One can see that the trick does not work if we have
the formula like:
$[\one A, \two B ] = \two X \one Y P.$ In this case
$$[\one A^n, \two B ]=
\sum_{i=0}^n \one A^i \two X \one Y  \two A^{n-1-i} P$$
and one is unable to calculate the trace of such expression
in the simple form.

So in order to calculate the
$[L^n(z), L^2(u) ] $ (as explained above)
we need to find the formula for the
$[L^n(z), L^2(u) ] $ in such form that
it is the sum of the type $\one A \two B P$.
It's done in the following lemma.

{\Lem \label{l-l2-hard}
\bea
[ \one L(z), \two L^2(u) ] &=&
(\one L^2(u)  - \two L^2(u) ) \frac{P}{z-u} \nn \\
&+&  (\one L(u)  + \two L(u) ) \two L(z) \frac{P}{z-u}
-\one L(z) (\one L(u) + \two L(u))  \frac{P}{z-u}
\eea
}

{\Lem From the lemma above one obtains
the following formula:
\label{ln-l2}
\bea
[ \one L^n(z), \two L^2(u) ] &=&
(\one L(u)  + \two L(u) ) \two L^n(z) \frac{P}{z-u}
-\one L^n(z) (\one L(u) + \two L(u))  \frac{P}{z-u}+ \nn \\ ~~
&+&\sum_{i=0}^{n-1}
\one L^i(z) ( \one L^2(u)   - \two L^2(u) ) \two L^{n-1-i}(z) \frac{P}{z-u} \nn \\ ~~
\eea
}
\\
Our next aim is to calculate
$[L^n(z), L^3(u) ] $.
As it was  explained in our receipt
we need to find the formula for the
$[L(z), L^3(u) ] $ in such form that
it is the sum of the type $\one A \two B P$.
It's done in the following lemma.

{\Lem \label{l-l3-hard}
\bea
[ \one L(z), \two L^3(u) ] &=&
(\one L^3(u)  - \two L^3(u) ) \frac{P}{z-u} \nn \\
&+&  (\one L^2(u)+ \one L(u) \two L(u)  + \two L^2(u) ) \two L(z)
\frac{P}{z-u} \nn \\
&-&\one L(z) (\one L^2(u)+ \one L(u) \two L(u)  + \two L^2(u) )  \frac{P}{z-u} \nn \\
&+& [\frac{\partial}{\partial u} \one L(u) +\frac{\one L(u)}{z-u}, \one
L(z)] \frac{1}{z-u}
\eea
}
\\
Going ahead let us note that as a corollary of the lemma above we see that
$ Tr [ L^2(u), L(z)]=0$.

{\Lem \label{ln-l3} From the lemma above one also sees that:
\bea
[ \one L^n(z), \two L^3(u) ] &=&
\sum_{i=0}^{n-1}
\one L^i(z) ( \one L^3(u)   - \two L^3(u) ) \two L^{n-1-i}(z) \frac{P}{z-u}
\nn \\ ~~
&+&  (\one L^2(u)+ \one L(u) \two L(u)  + \two L^2(u) ) \two L^n(z)
\frac{P}{z-u} \nn \\
&-&\one L^n(z) (\one L^2(u)+ \one L(u) \two L(u)  + \two L^2(u) )  \frac{P}{z-u} \nn \\
&+& [\frac{\partial}{\partial u} \one L(u) +\frac{\one L(u)}{z-u}, \one
L^n(z)] \frac{1}{z-u}
\eea
}

{\Rem \label{ln-lm} In general one
can guess the following sort of the
formula:
\bea
[ \one L^n(z), \two L^m(u) ] &=&
\sum_{i=0}^{n-1}
\one L^i(z) ( \one L^m(u)   - \two L^m(u) ) \two L^{n-1-i}(z) \frac{P}{z-u}
\nn \\
&+& ( \sum_{i=0}^{m-1} \one L^i(u)\two L^{m-1-i}(u) ) \two L^n(z) \frac{P}{z-u}
\nn \\
&-& \one L^n(z) ( \sum_{i=0}^{m-1} \one L^i(u)\two L^{m-1-i}(u) )
\frac{P}{z-u}\nn\\
&+& \mbox{``unwanted terms of lower degree''}
\eea
}
\\
But even in the case of m=4 it seems to be quite
difficult to obtain such formula.


\subsection{Commutativity of traces $[Tr L^n(z), Tr L^m(u)]=0$,
$n,m\le 3$ }

We will prove that $[Tr L^n(z), Tr L^m(u)]=0$, for $n\le 3, m\le 3$,  and
speculate about the general case. The case $n=2=m$ is easy and quite
well-known. In the next section we will prove that $[Tr L^4(z), Tr L^2(u)]
\ne 0$.

{\Lem \label{tr-ln-tr-l2}
The equality $ [ Tr  L^n(z), Tr L^2(u)] = 0$ is equivalent to
the equality $  Tr  [ L^n(z),  L(u)] = 0$,
moreover
$ [ Tr  L^n(z), Tr L^2(u)] = 2 Tr [L(u),  L^n(z) ] \frac{1}{z-u}$.
}
\\
{\bf Proof~}
\bea
&&[ Tr  L^n(z), Tr L^2(u)] =
 Tr [ \one L^n(z), \two L^2(u)] = \mbox{ / by lemma \ref{ln-l2} /}  \nn\\
&=& Tr \Bigl(
(\one L(u)  + \two L(u) ) \two L^n(z) \frac{P}{z-u}
-\one L^n(z) (\one L(u) + \two L(u))  \frac{P}{z-u} \nn \\
&+&\sum_{i=0}^{n-1}
\one L^i(z) ( \one L^2(u)   - \two L^2(u) ) \two L^{n-1-i} (z)
\Bigr)=  \mbox {by the formula \ref{Tr1} }\nn\\
&=& 2 Tr (L(u)  L^n(z) ) \frac{1}{z-u}
-2  Tr (L^n(z) L(u)) \frac{1}{z-u}+ \nn\\
&+&\sum_{i=0}^{n-1} Tr
L^i(z) L^2(u) L^{n-1-i} (z) - Tr  L^i(z) L^2(u) L^{n-1-i} (z) \nn \\
&=& 2 Tr [L(u) , L^n(z)] \frac{1}{z-u}
\eea
$\Box$

{\Lem \label{tr-ln-l}
The following equality holds:
\bea
&&Tr [L^n(z), L(u)]= \frac{1}{z-u}
\sum_{i=1}^{n-1}
\Bigl(
Tr L^i(z) L(u) Tr L^{n-1-i}(z) -
  Tr L^{n-1-i}(z) Tr L(u) L^i(z) \Bigr)
\nn\\
&&=\frac{1}{z-u}
\Bigl(
\sum_{i=1}^{n-1}
 [
Tr L^i(z) L(u),  Tr L^{n-1-i}(z)]
 - \sum_{i=0}^{n-2} Tr L^{i}(z) Tr [L(u), L^{n-1-i}(z)] \Bigr)
\eea
}
\\
{\bf Proof~} This is a straightforward application of
formula \ref{fml-ln-l-easy}
and formulas $Tr \one A \two B= Tr A Tr B \mbox{~~ and~~}
Tr[A,B]= Tr [\one A \two B] P:$
\bea
&&Tr [L^n(z), L(u)]=  Tr [\one L^n(z), \two L(u)] P
= \mbox{ by the formula \ref{fml-ln-l-easy} } \nn\\
&=&
Tr \Bigl(
(\two L^n(z)  - \one L^n(z) ) + 
\sum_{i=0}^{n-1} \one L^i(z) (\one L(u) - \two L(u)) \two L^{n-1-i}(z)
\Bigr) \frac{1}{z-u}  \nn \\
&=&
\Bigl(
(Tr Id) (Tr L^n(z)  - Tr L^n(z) ) \nn \\
&+&\sum_{i=0}^{n-1} Tr L^i(z)  L(u) Tr L^{n-1-i}(z)
 -  Tr L^i(z) Tr L(u)  L^{n-1-i}(z)
\Bigr) \frac{1}{z-u} \nn \\
&=&
\Bigl(
\sum_{i=0}^{n-1} Tr L^i(z)  L(u) Tr L^{n-1-i}(z)
 -  \sum_{i=0}^{n-1}  Tr L^{n-1-i}(z) Tr L(u)  L^{i}(z)
\Bigr) \frac{1}{z-u}  \nn \\
&& \mbox{ when i=0 then }
Tr  L(u) Tr L^{n-1}(z)
 -  Tr L^{n-1}(z) Tr L(u) = 0 \nn \\
&&
\mbox{ due to $Tr L(z)$ lies in the center of $\g(n)\oplus...\oplus \g(n)$ }
\nn \\
&=&
\Bigl(
\sum_{i=1}^{n-1} Tr L^i(z)  L(u) Tr L^{n-1-i}(z)
 -   Tr L^{n-1-i}(z) Tr L(u)  L^{i}(z)
\Bigr) \frac{1}{z-u} \nn \\
&=&
\frac{1}{z-u}
\Bigl(
\sum_{i=1}^{n-1}
 [
Tr L^i(z) L(u),  Tr L^{n-1-i}(z)]
 - \sum_{i=1}^{n-1}
  Tr L^{n-1-i}(z) Tr [L(u), L^i(z)] \Bigr)  \nn \\
&=&
\frac{1}{z-u}
\Bigl(
\sum_{i=1}^{n-1}
 [
Tr L^i(z) L(u),  Tr L^{n-1-i}(z)]
 -
\sum_{i=0}^{n-2}  Tr L^{i}(z) Tr [L(u), L^{n-1-i}(z)] \Bigr)
\eea
$\Box$

{\Cor \label{tr-l1-3-l1}
\bea
 Tr  [L^n(z),  L(u)] = 0 \mbox{ for n=1,2,3 }
\eea
}
\\
{\bf Proof~}
\\
Let $n=1$ - then by the lemma above we immediately
obtain zero, because the summation index $i$ is out of range
in the sum $\sum_{i=1}^{n-1}$.
\\
Let $n=2$, so  by the lemma above
\bea
&&(z-u) Tr  [L^2(z),  L(u)] = \sum_{i=1}^{1}
Tr L^i(z)  L(u) Tr L^{n-1-i}(z)
 -   Tr L^{n-1-i}(z) Tr L(u)  L^{i}(z)\nn\\
&=& (Tr Id ) Tr [L(z) , L(u)] = 0
\quad \mbox{ in virtue of the corollary in the case $n=1$ } \nn
\eea
\\
Let $n=3$, so  by the lemma above:
\bea
(z-u) Tr  [L^3(z),  L(u)] &=& \sum_{i=1}^{2}
Tr L^i(z)  L(u) Tr L^{n-1-i}(z)
 -   Tr L^{n-1-i}(z) Tr L(u)  L^{i}(z) \nn\\
&=& \bigl( Tr L(z)  L(u) Tr L(z)
 -   Tr L(z) Tr L(u)  L(z) \bigr)  \nn\\
&+& \bigl( Tr L^2(z)  L(u) (Tr (Id) )
 -   (Tr (Id) ) Tr L(u)  L^{i}(z) \bigr) = 0 \nn\\
&&\mbox{ in virtue of the collorary in the case $n=1,~n=2$
} \nn\\
&&\mbox{and the commutativity of $Tr L(z)$ with everything. } \nn
\eea
$\Box$

{\Cor \label{tr-l1-3-l1-n2}
One can simplify (reducing the range of summation)
 the expression
in lemma \ref{tr-ln-l} in the following way:
}
\bea
Tr [L^n(z), L(u)]
&=&\frac{1}{z-u}
\Bigl(
\sum_{i=1}^{n-3}
 [
Tr L^i(z) L(u),  Tr L^{n-1-i}(z)]\nn \\
& - & \sum_{i=0}^{n-5} Tr L^{i}(z) Tr [L(u), L^{n-1-i}(z)] \Bigr)
\eea

{\Cor \label{tr-l4-l1}
\bea
 Tr  [L^4(z),  L(u)] = \frac{1}{z-u} [Tr L(z)  L(u), Tr L^2(z)]
\eea
}

{\Cor \label{tr-l1-3-tr-l2}
\bea
[ Tr  L^n(z), Tr L^2(u)] = 0 \mbox{ for $n=1,2,3$ }
\eea
}
\\
{\bf Proof~} This follows from the corollary (\ref{tr-l1-3-l1}) and  lemma \ref{tr-ln-tr-l2}.
$\Box$

{\Lem \label{tr-ln-tr-l3}
The equality $ [ Tr  L^n(z), Tr L^3(u)] = 0$
follows from the equalities
 $  Tr  [ L^n(z),  L^2(u)] = 0$ and
 $  Tr  [ L^n(z),  L(u)] = 0$,
moreover
\bea \label{fml-tr-ln-tr-l3}
&&[ Tr  L^n(z), Tr L^3(u)] =  \nn\\
&& =\Bigl(
3  Tr  [L^2(u), L^n(z)]
\frac{1}{z-u}
+  (Tr Id) Tr [\frac{\partial}{\partial u}  L(u) +\frac{ L(u)}{z-u},
L^n(z)] \frac{1}{z-u}  \Bigr).
\eea
}
\\
{\bf Proof~}
\bea
&&[ Tr  L^n(z), Tr L^3(u)] =
 Tr [ \one L^n(z), \two L^3(u)] = \mbox{/ by lemma \ref{ln-l3} /}  \nn\\
&=& Tr \Bigl(
\sum_{i=0}^{n-1}
\one L^i(z) ( \one L^3(u)   - \two L^3(u) ) \two L^{n-1-i}(z) \frac{P}{z-u}
 \nn \\
&+& (\one L^2(u)+ \one L(u) \two L(u)  + \two L^2(u) ) \two L^n(z)
\frac{P}{z-u} \nn \\
&-&\one L^n(z) (\one L^2(u)+ \one L(u) \two L(u)  + \two L^2(u) )  \frac{P}{z-u} \nn \\
&+& [\frac{\partial}{\partial u} \one L(u) +\frac{\one L(u)}{z-u}, \one
L^n(z)] \frac{1}{z-u}  \Bigr) \nn\\
&=&
 \Bigl(
3  (Tr  L^2(u) L^n(z) \frac{1}{z-u} - Tr  L^n(z)  L^2(u) \frac{1}{z-u})+
\nn \\  &+&  (Tr Id) Tr [\frac{\partial}{\partial u}  L(u) +\frac{
L(u)}{z-u},
L^n(z)] \frac{1}{z-u}  \Bigr) \nn \\
&=&
 \Bigl(
3  Tr  [L^2(u), L^n(z)]
\frac{1}{z-u}
+  (Tr Id) Tr [\frac{\partial}{\partial u}  L(u) +\frac{ L(u)}{z-u},
L^n(z)] \frac{1}{z-u}  \Bigr) \nn
\eea
$\Box$

{\Cor \label{tr-l2-l2}
The following is true:
\bea
Tr [L^2(z), L^2(u)]=0
\eea
}
\\
{\bf Proof~}
This follows from corollary \ref{tr-l1-3-tr-l2}
that $ [Tr L^2(z), Tr L^3(u)]=0$ and from  corollary \ref{tr-l1-3-l1}
that  $ Tr [L(u), L^2(z) ] =0 $, so according to the lemma above
formula \ref{fml-tr-ln-tr-l3}
applied to the case $n=2$
gives that:
 $$ 0 = \frac{3}{z-u}[Tr L^2(u), Tr L^2(z)] +0 $$
$\Box$

{\Lem \label{tr-ln-l2}
The following equality holds:
\bea &(z-u)Tr [   L^n(z), L^2(u)] = &\nn\\
&= Tr [L(u),  L^n(z)]+
\sum_{i=0}^{n-3}
[Tr L^i(z) L^2(u), Tr L^{n-1-i}(z) ] -
Tr L^{i}(z) Tr [L^2(u), L^{n-1-i}(z)]
&\nn
\eea
}

{\bf Proof~} \bea
&& Tr [   L^n(z), L^2(u)] = Tr [   \one L^n(z), \two L^2(u)] P =
\mbox{/ by lemma \ref{ln-l2} /} \nn\\
&=&
Tr \Bigl((\one L(u)  + \two L(u) ) \two L^n(z) \frac{1}{z-u}
-\one L^n(z) (\one L(u) + \two L(u))  \frac{1}{z-u} \nn \\
&+&\sum_{i=0}^{n-1}
\one L^i(z) ( \one L^2(u)   - \two L^2(u) ) \two L^{n-1-i}(z)
\frac{1}{z-u} \Bigr) \nn \\
&=&
\frac{1}{z-u} \Bigl( [Tr L(u), Tr L^n(z)]+Tr [L(u),  L^n(z)] \nn \\ ~~
&+&\sum_{i=0}^{n-1}
Tr L^i(z) L^2(u) Tr L^{n-1-i}(z) -
Tr L^i(z) Tr L^2(u) L^{n-1-i}(z)
\Bigr) \nn \\
&=&
\frac{1}{z-u} \Bigl( Tr [L(u),  L^n(z)]+
\sum_{i=0}^{n-1}
Tr L^i(z) L^2(u) Tr L^{n-1-i}(z) \nn\\
&-& Tr L^{n-1-i}(z) Tr L^2(u) L^{i}(z)
\Bigr) \nn \\
&=&
\frac{1}{z-u} \Bigl( Tr [L(u),  L^n(z)]+
\sum_{i=0}^{n-1}
[Tr L^i(z) L^2(u), Tr L^{n-1-i}(z) ] \nn\\
&-& Tr L^{n-1-i}(z) Tr [L^2(u), L^{i}(z)]
\Bigr)
\mbox { by corollaries \ref{tr-l1-3-l1}, \ref{tr-l2-l2} } \nn\\
&=&
\frac{1}{z-u} \Bigl( Tr [L(u),  L^n(z)]+
\sum_{i=0}^{n-3}
[Tr L^i(z) L^2(u), Tr L^{n-1-i}(z) ] \nn\\
&-&Tr L^{i}(z) Tr [L^2(u), L^{n-1-i}(z)]
\Bigr) 
\eea
$\Box$

{\Cor \label{tr-l3-l2} The following is true: \bea Tr [L^3(z), L^2(u)]=0
\eea }
\\
{\bf Proof~} According to the lemma above we obtain:
\bea && (z-u) Tr [L^3(z),
L^2(u)]\nn\\
&=& Tr [L(u),  L^3(z)]+ \sum_{i=0}^{0} [Tr L^i(z) L^2(u), Tr
L^{2-i}(z) ] - Tr L^{i}(z) Tr [L^2(u), L^{2-i}(z)] \nn\\
&=& Tr [L(u),  L^3(z)]+
[Tr L^2(u), Tr L^{2}(z) ] - Tr Id  Tr [L^2(u), L^{2}(z)]=0
\eea in virtue of
corollaries \ref{tr-l1-3-l1}, \ref{tr-l1-3-tr-l2}, \ref{tr-l2-l2}. $\Box$

{\Cor \label{tr-l3-tr-l3}
The following is true:
\bea
[Tr L^3(z), Tr L^3(u)]=0
\eea
}
\\
{\bf Proof~} this is directly implied by lemma
\ref{tr-ln-tr-l3}
because the conditions requested in this lemma
are true by corollaries \ref{tr-l1-3-l1}, \ref{tr-l3-l2}
$\Box$


\subsection{NONcommutativity of traces $L^4, L^2 $}

According to lemma \ref{tr-ln-tr-l2}
and corollary \ref{tr-l4-l1}
\bea
[ Tr  L^4(z), Tr L^2(u)] =
\frac{2}{z-u} Tr [L(u),  L^4(z) ]
=
\frac{-2}{(z-u)^2} [Tr L(z)  L(u), Tr L^2(z)]
\eea
\\
Let us consider for simplicity the
case $\K=0$.

\bea
Tr L^2(z) =
\sum_i \frac{Tr \Phi_i^2}{(z-a_i)^2}
+\sum_k \frac{1}{z-a_k} \sum_{j\ne k} \frac{2 Tr \Phi_k \Phi_j}{(a_k-a_j)}
\eea

\bea
H_k=\sum_{j\ne k} \frac{2 Tr \Phi_k \Phi_j}{(a_k-a_j)}
\eea
\\
It follows from $ [Tr L^2(u), Tr L^2(z)]=0 $
that $[H_k, H_j]=0$.

\bea Tr L(z) L(u)  = \sum_i \frac{Tr \Phi_i^2}{(z-a_i)(u-a_i)} +\sum_m
\frac{1}{u-a_m} \sum_{n\ne m} \frac{ Tr \Phi_m \Phi_n}{z-a_n} \eea
\\
The commutator
$[Tr L(z)  L(u), Tr L^2(z)]$ vanishes if
\bea
0=[\sum_m \frac{1}{z-a_m} \sum_{n\ne m} \frac{ Tr \Phi_m \Phi_n}{u-a_n},
\sum_k \frac{1}{u-a_k} \sum_{j\ne k} \frac{2 Tr \Phi_k \Phi_j}{(a_k-a_j)}]
\eea
\\
This expression is a rational function on variables $z,u.$ It is zero iff
its residues on $z$ are zeros. For example, its residue at $z=a_m$ is a
rational function on $u$ with double poles at $u=a_i,$ let us consider its
coefficient at $\frac{1}{(u-a_n)^2}$
\bea [  Tr \Phi_m
\Phi_n,
 \sum_{j\ne n} \frac{2 Tr \Phi_n \Phi_j}{(a_n-a_j)}]
\eea
This could be zero for any $a_k$ only if
the expression below  should be zero for all $m,n,j\ne m$
\bea
&&[  Tr \Phi_m \Phi_n,
 Tr \Phi_n \Phi_j]
= Tr
[  \one  \Phi_m  \one \Phi_n,
 \two  \Phi_n \two \Phi_j]
=
Tr
  \one  \Phi_m  [\one \Phi_n,
 \two  \Phi_n]  \two \Phi_j
=
Tr
  \one  \Phi_m  [\one \Phi_n, P]  \two \Phi_j
\nn\\
&&=
Tr
  \one  \Phi_m  \one \Phi_n  \one \Phi_j P
- Tr P \two  \Phi_m  \one \Phi_n  \two \Phi_j
=
Tr
  \one  \Phi_m  \one \Phi_n  \one \Phi_j P
- Tr P \two  \Phi_m   \two \Phi_j \one \Phi_n
\nn\\
&&= Tr \Phi_m  \Phi_n  \Phi_j -
Tr  \Phi_m    \Phi_j \Phi_n
\neq \mbox{ZERO !!!}
\eea
Let us note that there is no such a problem
for the case of one- or two-poles Gaudin system.

So we obtain that
\bea
 0\neq [ Tr  L^4(z), Tr L^2(u)] =
\frac{2}{z-u} Tr [L(u),  L^4(z) ]
=
\frac{-2}{(z-u)^2} [Tr L(z)  L(u), Tr L^2(z)]
\eea
$\Box$


\subsection{One spin Gaudin model (argument shift method)}

The one spin Gaudin's model Lax  operator is the following one:
\bea
L_{MF}(z)= \K + \frac{\Phi}{z}
\eea
\\
It's well-known that considering the coefficients at ${z}^{-1}$
of $\s(Tr L_{MF}^k(z)) \in S(\g(n))$ one obtains the maximal Poisson-commutative
subalgebra in $S(\g(n))$ (for general $\K$) \cite{Mishen-Fomen,Vinberg}. The
restriction of such subalgebra to a generic coadjoint orbit gives a
classical integrable system.

The aim of this section is to discuss the lifting of this maximal
commutative subalgebra from $S(\g(n))$ to $U(\g(n))$, by considering
$Tr L_{MF}^k(z) \in U(\g(n))$. In other words we discuss
the quantum integrability of the one spin Gaudin model. We prove
commutativity in the following restricted case:
$$[ Tr L_{MF}^k(z)) , Tr L_{MF}^l(u))]=0\qquad \mbox{for} \qquad k,l\le4$$
this is more optimistic than the general Gaudin model, however
we also prove that $$ [ Tr L_{MF}^3(z)) , Tr L_{MF}^6(u))]\ne 0.$$

\subsubsection{Preliminary remarks}

{\Lem \label{mf-l-l} The following  holds: \bea [  L_{MF}(u),
L_{MF}(z)]=[\K, \Phi] (\frac{1}{u}-\frac{1}{z}) =[\K,
L_{MF}(z)](\frac{z}{u}-1) \eea }

{\Rem \bea L_{MF}(u)=\frac{z}{u} L_{MF}(z) - K(\frac{z}{u}-1) ~~~~~~~
L_{MF}(z)= K(1-\frac{u}{z}) +\frac{u}{z} L_{MF}(u) \eea }
\\
Let $A$ be a linear operator in $V\otimes V.$ Let us denote $Tr_1 A$
the trace taken  only over the first component,
i.e. $(Tr_1 A)_{ij}= \sum_k A_{kk,ij}$.
\\
By a straightforward calculation one has
{\Lem \label{tr1-A1P-ii}
\bea
Tr_1(\one A P)=Tr_1(\two A P)=A\qquad Tr_1(\one A \two B P)=\sum_{ij}\sum_k A_{kj}B_{ik}
e_{ij}\nn
\eea
}

\subsubsection{Quadratic Hamiltonians}

{\Lem \label{mf-tr-ln-tr-l2} The following holds:
\bea [ Tr L_{MF}^n(z),
Tr  L_{MF}^2(u)]=Tr [L_{MF}(u),  L_{MF}^n(z) ]=0 \eea } {\bf Proof~}
\bea &&[ Tr
L_{MF}^n(z), Tr  L_{MF}^2(u)]=
\mbox{ according to lemma \ref{tr-ln-tr-l2}}\nn\\
&=& 2 Tr [L_{MF}(u),  L_{MF}^n(z) ] \frac{1}{z-u}=
Tr \frac{2}{z-u} \sum_{i=0}^{n-1} L_{MF}^i(z) [L_{MF}(u),  L_{MF}(z) ] L_{MF}^{n-i-1}(z) \nn\\
&=& Tr\frac{2}{z-u} \sum_{i=0}^{n-1} L_{MF}^i(z) [\K,
L_{MF}(z)](\frac{z}{u}-1) L_{MF}^{n-i-1}(z)
 \nn\\
&=&
 \frac{2}{z-u}(\frac{z}{u}-1)  Tr [\K, L_{MF}^n(z)]=0
\mbox{ because $\K$ is a $\CC$-valued matrix}
\eea
$\Box$
\\

\subsubsection{Cartan subalgebra}

We will prove that Cartan subalgebra is contained is
the subalgebra in $U(\g(n))$ generated by the $Tr (L(z)^k)$ and
Cartan subalgebra commutes with any element of type $Tr (L(z)^k)$.

{\Lem \label{Cartan} Let us assume that $\K=diag \{ k_j \} $, then the Cartan
subalgebra generated by $e_{jj}$ is contained in the subalgebra generated by
coefficients in $z$ of $ Tr L_{MF}(z)^k $. }
\\
{\bf Proof~} Let us consider the residue of $Tr L_{MF}(z)^k$ at ${z}=0$
$$Res_{z=0}Tr L^k_{MF}(z)= Res_{z=0}Tr \left(\K + \frac{\Phi}{z}\right)^k=k
Tr \K^{k-1} \Phi =\sum_j k_j^{k-1} e_{jj}. $$
The general choice of $\K$ imply that all the $k_j$ are different. In this case
the Vandermonde matrix $\{k_j^l\}$ is not degenerate and by taking  linear combinations
of the residues above one recovers $ e_{jj} $.

{\Lem \label{mf-tr-ln-ljj}  $ \forall n,i$ one has
$$[ Tr L_{MF}^n(z), (L_{MF}(u))_{ii}] = [ Tr  L_{MF}^n(z), e_{ii}] = 0.$$}
\\
{\bf Proof~}
\bea
&& [ Tr  L_{MF}^n(z), (L_{MF}(u))_{ii}] = \sum_{j}   [  \one L_{MF}^n(z),
\two (L_{MF}(u))]_{jj,ii} \nn\\
&=& \Bigl( Tr_{1} [  \one L_{MF}^n(z), \two
(L_{MF}(u))] \Bigr)_{ii} = \mbox{ /by the formula \ref{fml-ln-l-easy}/ }\nn\\
&=&\Bigl( Tr_{1}\Bigl( (\two L_{MF}^n(z)  - \one L_{MF}^n(z) ) \frac{P}{z-u}+
\sum_{k=0}^{n-1} \one L_{MF}^k(z) (\one L_{MF}(u) - \two L_{MF}(u)) \two
L_{MF}^{n-1-k}(z) \frac{P}{z-u} \Bigr)\Bigr)_{ii}\nn
\eea
\bea
&=&\mbox{ /by lemma \ref{tr1-A1P-ii} /}
= ( L_{MF}^n(z))  - (L_{MF}^n(z) )_{ii} \frac{1}{z-u}\nn\\
&+&
\frac{1}{z-u}  \sum_{k=0}^{n-1}  \sum_j ( L_{MF}^k(z) L_{MF}(u))_{ji}
(L_{MF}^{n-1-k}(z))_{ij} - ( L_{MF}^k(z))_{ji} (L_{MF}(u)
L_{MF}^{n-1-k}(z))_{ij} \nn\\
&=& \frac{1}{z-u}  \sum_{k=0}^{n-1}  \sum_j (
L_{MF}^k(z) (\frac{z}{u} L_{MF}(z) - \K (\frac{z}{u}-1)) )_{ji}
(L_{MF}^{n-1-k}(z))_{ij}\nn\\
&-& ( L_{MF}^k(z))_{ji} (\frac{z}{u} L_{MF}(z) -
\K (\frac{z}{u}-1) L_{MF}^{n-1-k}(z))_{ij}
\nn\\
&=& \frac{1}{z-u} \frac{z}{u}
\sum_{k=0}^{n-1}  \sum_j ( L_{MF}^k(z) L_{MF}(z) )_{ji}
(L_{MF}^{n-1-k}(z))_{ij} \nn\\
&-&
( L_{MF}^k(z))_{ji} (L_{MF}(z)
L_{MF}^{n-1-k}(z))_{ij} - \frac{1}{z-u}  (\frac{z}{u}-1) \sum_{k=0}^{n-1}
\sum_j ( L_{MF}^k(z) \K)_{ji} (L_{MF}^{n-1-k}(z))_{ij} \nn\\
&-& (L_{MF}^k(z))_{ji} (\K L_{MF}^{n-1-k}(z))_{ij}
=\frac{1}{z-u} \frac{z}{u}    \sum_j ( L_{MF}^{n}(z) )_{ji} \delta_{ij}
 - \delta_{ij}
( L_{MF}^{n}(z))_{ij} \nn\\
 &-& \frac{1}{z-u}
(\frac{z}{u}-1) \sum_{k=0}^{n-1}  \sum_j (  L_{MF}^k(z)_{ji} k_i
L_{MF}^{n-1-k}(z)_{ij} -
 L_{MF}^k(z)_{ji}  k_i
L_{MF}^{n-1-k}(z)_{ij}) = 0 \nn
\eea $\Box$

\subsubsection{Commutativity of traces $ [ Tr  L_{MF}^n(z), Tr
L_{MF}^3(u)] = 0$ for n=1,...,5 }

{\Lem \label{mf-tr-ln-tr-l3}
\bea
 [ Tr  L_{MF}^n(z), Tr
L_{MF}^3(u)] =
\frac{3}{u^2}
 \sum_{i=3}^{n-2} ( [ Tr
L_{MF}^{i}(z),    Tr  \K^2    L_{MF}^{n-1-i}(z) ])
\label{fml-mf-tr-ln-tr-l3}
\eea
}

{\Cor
\bea
 [ Tr  L_{MF}^n(z), Tr
L_{MF}^3(u)] = 0, \mbox{ for n=1,...,5}
\eea
}
\\
{\bf Proof of the Corollary~} The corollary follows immediately
for $n=1,...,4$ from the lemma above - because
summation in the formula \ref{fml-mf-tr-ln-tr-l3}
 is out of range
for such $n$. The only case is $n=5$
then there is only one term in summation:
$[ Tr
L_{MF}^{3}(z),    Tr  \K^2    L_{MF}(z) ] $,
this term equals to zero due to the lemma
\ref{mf-tr-ln-ljj}.
\\
{\bf Proof of the lemma~} According to
the lemma \ref{mf-tr-ln-tr-l2} one has
$  Tr  [ L_{MF}^n(z),  L_{MF}(u)] = 0.$
Hence using the  lemma \ref{tr-ln-tr-l3},
we obtain
\bea
 [ Tr  L_{MF}^n(z), Tr
L_{MF}^3(u)] = \frac{3}{z-u}
Tr  [L_{MF}^2(u), L_{MF}^n(z)]
\eea
Let us simplify this expression:
\bea && Tr [   L_{MF}^n(z), L_{MF}^2(u)] =
Tr[ L_{MF}^n(z), (\frac{z}{u} L_{MF}(z) - K(\frac{z}{u}-1))^2]\nn\\
&=&
-(\frac{z}{u}-1)\frac{z}{u} Tr[ L_{MF}^n(z), ( K L_{MF}(z) + L_{MF}(z)K)]\nn\\
&=&
-(\frac{z}{u}-1)\frac{z}{u} Tr L_{MF}^n(z) K L_{MF}(z) - L_{MF}(z)K L_{MF}^n(z)\nn
\eea
\\
So we came to the expression which is considered in the lemma
\ref{l-BKA-AKB}.
Let us denote
$$B= L_{MF}^n(z), \qquad A=L_{MF}(z).$$
Then using $$ [ \one L_{MF}(z) , \two
L_{MF}(z)]=\frac{1}{z}[\one L_{MF}(z)- \one \K,P]= (\one L_{MF}(z) - \one
\K -  \two L_{MF}(z) + \two \K) P $$ we obtain
\bea
 [ \one B , \two A] &=& [ \one L_{MF}^n(z) , \two L_{MF}(z)] =
\nn\\
&=&
\sum_{i=0}^{n-1} \frac{1}{z} \one L_{MF}^{i}(z) ( \one L_{MF}(z) - \one \K
-   \two L_{MF}(z) + \two \K  ) \two L_{MF}^{n-1-i}(z) P =
\nn\\
&=&
\frac{1}{z} (
\one L_{MF}^{n}(z) -\two L_{MF}^{n}(z))P + \sum_{i=0}^{n-1} \frac{1}{z}
\one L_{MF}^{i}(z) ( - \one \K + \two \K  ) \two L_{MF}^{n-1-i}(z) P.
\eea
\\
As the result we get
\bea
&& Tr (P \two \K + \two \K P) ([ \one B , \two A] )
=
Tr (P\two K
+ \two K P) [ \one L_{MF}^n(z) , \two L_{MF}(z)]
\nn\\
&=&
Tr (P\two K + \two K
P)( \frac{1}{z} ( \one L_{MF}^{n}(z) -\two L_{MF}^{n}(z))P +
\sum_{i=0}^{n-1} \frac{1}{z} \one L_{MF}^{i}(z) ( - \one \K + \two \K  )
\two L_{MF}^{n-1-i}(z) P)
\nn\\
&=&
\frac{1}{z} Tr (\two K \one L_{MF}^{n}(z) +
\one \K \one L_{MF}^{n}(z) - \two \K \two L_{MF}^{n}(z) -\one \K \two
L_{MF}^{n}(z))
\nn\\
&+&
\frac{1}{z} Tr  (\two K + \one K ) \sum_{i=0}^{n-1}  \one
L_{MF}^{i}(z) ( - \one \K + \two \K  ) \two L_{MF}^{n-1-i}(z)
\nn\\
&=&
\frac{1}{z} Tr (\two K \one L_{MF}^{n}(z) + \one \K \one L_{MF}^{n}(z) -
\two \K \two L_{MF}^{n}(z) -\one \K \two L_{MF}^{n}(z))
\nn\\
&+&
\frac{1}{z} Tr
(\two K + \one K ) \sum_{i=0}^{n-1}  \one L_{MF}^{i}(z) ( - \one \K + \two
\K  ) \two L_{MF}^{n-1-i}(z)
\nn\\
&=&
\frac{1}{z}  \sum_{i=0}^{n-1} Tr (- \two \K
\one L_{MF}^{i}(z)   \one \K \two L_{MF}^{n-1-i}(z) +
\nn
\eea
\bea
&+&
\two \K  \one
L_{MF}^{i}(z)  \two \K   \two L_{MF}^{n-1-i}(z) - \one \K  \one
L_{MF}^{i}(z)  \one \K  \two L_{MF}^{n-1-i}(z) + \one \K  \one
L_{MF}^{i}(z)  \two \K   \two L_{MF}^{n-1-i}(z) )
\nn\\
&=&
\frac{1}{z}
\sum_{i=0}^{n-1} (-   Tr  L_{MF}^{i}(z)   \K  Tr  L_{MF}^{n-1-i}(z)  \K
\nn\\
&+&
Tr   L_{MF}^{i}(z)   Tr  \K    L_{MF}^{n-1-i}(z)  \K - Tr \K L_{MF}^{i}(z)
\K  Tr   L_{MF}^{n-1-i}(z) + Tr \K   L_{MF}^{i}(z)  Tr \K
L_{MF}^{n-1-i}(z) )
\nn\\
&=&
 \frac{1}{z}  \sum_{i=0}^{n-1} ( Tr   L_{MF}^{i}(z)
Tr  \K    L_{MF}^{n-1-i}(z)  \K - Tr \K   L_{MF}^{i}(z)  \K  Tr
L_{MF}^{n-1-i}(z) )
\nn\\
&=&
\frac{1}{z}  \sum_{i=0}^{n-1} ( [ Tr L_{MF}^{i}(z),
Tr  \K    L_{MF}^{n-1-i}(z)  \K ])=
\mbox{/using lemma \ref{TrKPhi-TrK2Phi} /}
\nn\\
&=& \frac{1}{z} \sum_{i=3}^{n-2} ( [ Tr
L_{MF}^{i}(z),    Tr  \K    L_{MF}^{n-1-i}(z) \K ])\nn
\eea
$\Box$

\subsubsection{NONcommutativity of traces of $L^6$ and $L^3$}

{\Lem $[Tr L_{MF}^6(z), Tr L_{MF}^3(z) ] \ne 0$}
\\
{\bf Proof~}
According to lemma \ref{mf-tr-ln-tr-l3} one has
\bea
[Tr L_{MF}^6(z), Tr L_{MF}^3(z) ]=
\frac{3}{z^2}
( [ Tr L_{MF}^{3}(z),    Tr  \K^2    L_{MF}^{2}(z) ])
+
\frac{3}{z^2}
( [ Tr L_{MF}^{4}(z),    Tr  \K^2    L_{MF}(z) ])
\eea
The second term equals to zero due to the lemma \ref{mf-tr-ln-ljj}.
Let us show that the first term is non zero.

\bea
&&[ Tr   L_{MF}^{3}(z),    Tr  \K^2    L_{MF}^{2}(z)  ]
\nn\\
&=&
 [ Tr   \frac{\K^2\Phi}{z}+ \frac{2\K\Phi^2 +\Phi\K\Phi}{z^2}
,    2 Tr  \frac{\K^3 \Phi}{z}  ]
+
 [ Tr   \frac{\K^2\Phi}{z}+ \frac{2\K\Phi^2 +\Phi\K\Phi}{z^2}
,    Tr  \frac{\K^2 \Phi^2}{z^2}  ]
\nn\\
&&\mbox{/ according to lemmas \ref{l-Skr}, \ref{faf-bf} the first commutator equals to
zero,}
\nn\\
&&\mbox{
 $[ Tr   \frac{\K^2\Phi}{z},  Tr  \frac{\K^2 \Phi^2}{z^2}]=0$ by lemma \ref{l-Skr}/
}
\nn\\
&=&
\frac{1}{z^4} [ Tr   2\K\Phi^2 +\Phi\K\Phi, Tr  \K^2 \Phi^2  ]
=
\frac{1}{z^4} Tr [  2 \one \K \one \Phi^2 +\one \Phi \one  \K \one\Phi,
 \two\K^2 \two \Phi^2  ]
\nn\\
&=&
\frac{1}{z^4} (2 Tr \one \K \two\K^2 [\one \Phi^2, \two \Phi^2  ]
+
Tr (\two\K^2 [\one \Phi, \two \Phi^2 ]\one  \K \one\Phi +
\one \Phi \one  \K  \two\K^2 [ \one\Phi , \two \Phi^2  ]))
\nn\\
&=&
\frac{1}{z^4} (2 Tr \one \K \two\K^2
(P\two \Phi ^3 - P \one \Phi ^3
+ P \one \Phi \two \Phi ^2 - P \one \Phi ^2 \two \Phi )
+
Tr (\two\K^2 [ P, \two \Phi^2 ]\one  \K \one\Phi +
\one \Phi \one  \K  \two\K^2 [ P , \two \Phi^2  ]))
\nn\\
&=&
\frac{1}{z^4} (
2
(  Tr \one \K \two\K^2 \one \Phi ^3 P - Tr \one \K \two\K^2  \two \Phi ^3 P
+Tr \one \K \two\K^2  \two \Phi \one \Phi ^2 P
- Tr \one \K \two\K^2  \two \Phi ^2 \one \Phi  P)
\nn\\
&+&
Tr (\two\K^2  P \two \Phi^2 \one  \K \one\Phi
- \two\K^2 \two \Phi^2 P \one  \K \one\Phi +
\one \Phi \one  \K  \two\K^2  P  \two \Phi^2
- \one \Phi \one  \K  \two\K^2 \two \Phi^2  P))
\nn\\
&=&
\frac{1}{z^4} (
2
(  Tr \K^3  \Phi ^3  - Tr \K^3  \Phi ^3
+ Tr \K^2  \Phi  \K  \Phi ^2
- Tr  \K^2  \Phi ^2  \K  \Phi  )
\nn
\eea
\bea
&+&
Tr (    \Phi^2   \K^3  \Phi
- \K^2  \Phi^2   \K \Phi +
 \Phi   \K   \Phi^2 \K^2
-  \Phi \K^3 \Phi^2 ))
\nn\\
&=&
\frac{1}{z^4} (
2
(Tr \K^2  \Phi  \K  \Phi ^2
- Tr  \K^2  \Phi ^2  \K  \Phi  )
+
Tr (
- \K^2  \Phi^2   \K \Phi +
 \Phi   \K   \Phi^2 \K^2
))
\nn\\
&=&
\frac{1}{z^4} 3 (
Tr \K^2  \Phi  \K  \Phi ^2
- Tr  \K^2  \Phi ^2  \K  \Phi  )\nn
\eea
\\
We have used $ Tr \Phi^n \K \Phi - \Phi \K \Phi^n = 0 $
for $n=3$. This will be proved latter
(see lemma \ref{phin-k-phi}).
\\
So we have just obtained that
$$ [ Tr   L^{3}(z),    Tr  \K^2    L^{2}(z)  ]=
\frac{3}{z^4}  (
 Tr \K^2  \Phi  \K  \Phi ^2
- Tr  \K^2  \Phi ^2  \K  \Phi  )
$$
and
$$[ Tr   \K\Phi^2  , Tr \K^2   \Phi ^2]
=
[ Tr \Phi\K\Phi  , Tr \K^2   \Phi ^2]
=
 Tr \K^2  \Phi  \K  \Phi ^2
- Tr  \K^2  \Phi ^2  \K  \Phi.
$$
\\
Let us show that
$ Tr \K^2  \Phi  \K  \Phi ^2
- Tr  \K^2  \Phi ^2  \K  \Phi \ne 0 $.
\\
Recall that we are considering the diagonal matrix
$\K = \{ k_j \}, $
then:
\bea
 Tr \K^2  \Phi  \K  \Phi ^2
- Tr  \K^2  \Phi ^2  \K  \Phi
&=&
\sum_{j,p} k_j^2 \Phi_{jp} k_p  (\Phi)^2_{pj}
- k_j^2 (\Phi)^2_{jp} k_p  (\Phi)_{pj}\nn\\
&=&
\sum_{j \ne p} k_j^2 k_p ( \Phi_{jp}  (\Phi)^2_{pj} - (\Phi)^2_{jp}
(\Phi)_{pj} ). \nn
\eea
For this expression to be zero it should be that $$\forall
j,p :   \Phi_{jp}  (\Phi)^2_{pj} - (\Phi)^2_{jp} \Phi_{pj} =0,$$ but it's
not true for  the case of $\g(3)$ and higher rank algebras, by the
PBW property, due to the following
$$ \Phi_{jp}  (\Phi)^2_{pj} - (\Phi)^2_{jp} \Phi_{pj}
=
\Phi_{jp}  \sum_l \Phi_{pl } \Phi_{lj } -
\sum_l \Phi_{jl} \Phi_{lp} \Phi_{pj}.
$$
Hence the term
$ \sum_l \Phi_{j,p}  \Phi_{p,l } \Phi_{l,j}$
contains elements $ \Phi_{p,l }$ which are not contained
in the term: $-
\sum_l \Phi_{j,l} \Phi_{l,p} \Phi_{p,j}
$.
$\Box$

{\Rem ~~} For  the case of $\g(2)$
this is  zero (we should check only  j=1,p=2):
\bea
\Phi_{1 2}   \Phi_{2 1 } \Phi_{1 1 } +
\Phi_{1 2}   \Phi_{2 2 } \Phi_{2 1 }
- \Phi_{1 1} \Phi_{1 2} \Phi_{2 1} -
\Phi_{1 2} \Phi_{2 2} \Phi_{2 1}
=
[\Phi_{1 2}   \Phi_{2 1 } ,  \Phi_{1 1 }]
\nn\\
=
[\Phi_{1 2} ,     \Phi_{1 1 }]\Phi_{2 1 }
+ \Phi_{1 2}  [  \Phi_{2 1 },   \Phi_{1 1 }]
=
- \Phi_{1 2}   \Phi_{2 1 } +
\Phi_{1 2}   \Phi_{2 1 }
= 0\nn
\eea


\subsubsection{Commutativity $[Tr L_{MF}^4(z),Tr L_{MF}^4(u) ]= 0 $ }

{\Lem ~~ \label{l-mf4} The following is true: \bea \label{fml-mf4} [Tr
L_{MF}^4(z),Tr L_{MF}^4(u) ]= 0. \eea }
\\
Combined with the results of the previous sections this lemma proves the
following

{\Th ~~ \label{th-mf4}  Coefficients at $\frac{1}{z^i}$ of $Tr
L_{MF}^l(z)$, $l\le N$ freely generates  maximal commutative subalgebra in
$U(\g(N))$,  $N\le4$. }
\\
Let us explain that the maximality and free generation follows from the
results of Mishchenko and Fomenko \cite{Mishen-Fomen}, who proved that the
coefficients at $\frac{1}{z}$ of $\s( Tr L_{MF}^k(z)) \in S(\g(n))$
generate maximal Poisson-commutative subalgebra in $S(\g(n))$ (for general
$\K$).

{\Rem ~~ } According to the previous results the theorem above cannot be true
for  $N\ge 6$. We hope that it is still true for $N=5$.
\\
{\bf Proof of the lemma \ref{l-mf4} }
\\
We do not have the general formula for the commutators $ [L_{MF}^n(z),
L_{MF}^4(u) ]$ and propose here a straightforward proof.

\bea &&[Tr L_{MF}^4(z), Tr L_{MF}^4(u) ]=
[Tr (\K+\frac{\Phi}{z})^4, Tr (\K+\frac{\Phi}{u})^4 ]\nn\\
%
&=&[Tr (3\K^2\Phi^2+2\K\Phi\K\Phi+\Phi \K^2 \Phi),
 2 Tr (\K\Phi^3+\Phi\K\Phi^2) ] ( \frac{1}{z^2u^3}-  \frac{1}{z^3u^2})
\nn\\
&=&[Tr (4\K^2\Phi^2+2\K\Phi\K\Phi+[\Phi, \K^2 \Phi]),
 2 Tr (2\K\Phi^3+[\Phi,\K\Phi^2]) ] ( \frac{1}{z^2u^3}-  \frac{1}{z^3u^2})
\nn\\
&=&\mbox{/ by lemma \ref{lem-k-phi-n} /}
=[Tr ([\Phi, \K^2 \Phi]),
 2 Tr (2\K\Phi^3+[\Phi,\K\Phi^2]) ] ( \frac{1}{z^2u^3}-  \frac{1}{z^3u^2})
\nn\\
&+&[Tr (4\K^2\Phi^2+2\K\Phi\K\Phi),
 2 Tr ([\Phi,\K\Phi^2]) ] ( \frac{1}{z^2u^3}-  \frac{1}{z^3u^2})
\nn\\
&=&\mbox{/ by lemma \ref{lem-phi-k-phi} and the fact that $Tr \Phi^k$ are Casimirs / }
\nn\\
&=&
[ -Tr (\K^2 \Phi)~~ Tr Id,
 2 Tr (2\K\Phi^3) -Tr (\K \Phi^2)~~ Tr Id ] ( \frac{1}{z^2u^3}-  \frac{1}{z^3u^2})
\nn\\
&+&[Tr (4\K^2\Phi^2+2\K\Phi\K\Phi),
  - 2 Tr (\K \Phi^2)~~ Tr Id ] ( \frac{1}{z^2u^3}-  \frac{1}{z^3u^2})
\nn\\
&&\mbox{/ which is zero due to the lemma \ref{lem-k-phi-n} and
 \ref{l-Skr}. /}
\nn
\eea
$\Box$

Analogously we can prove that \bea [Tr
(2\K^2\Phi^2+\K\Phi\K\Phi),  Tr (\Phi^l \K\Phi^n)]=0 \nn\\ ~~ [ Tr
(3\K^2\Phi^2+2\K\Phi\K\Phi+\Phi \K^2 \Phi), Tr (\Phi^l \K\Phi^n)]=0 \eea
hence the coefficient at $\frac{1}{z^2}$ of $Tr L_{MF}^4(z)$ commutes with
the coefficient at $\frac{1}{z^{l+n}}$ of $Tr L_{MF}^{l+n+1}(z)$.



\subsubsection{Auxiliary lemmas.}

In this subsection we will prove some lemmas,
which has been used in the calculations above,
but may also represent an independent interest.

{\Lem \label{l-BKA-AKB}
\bea \label{BKA-AKB}
Tr  B \K A- A\K B
=
Tr P (\two K + \one K )
([ \one B , \two A] )
\mbox{  for $A, B: [A,B]=0$ }
\eea
}
\\
{\bf Proof~}
Let us assume that $\K=diag \{k_j \}$.
Then
\bea
&&Tr  B \K A- A\K B= \sum_j k_j (\sum_i
B_{ij}A_{ji}-A_{ij}B_{ji})
\nn\\
&=&\sum_j k_j (\sum_i A_{ji} B_{ij}-B_{ji}
A_{ij}+ [B_{ij},A_{ji}]-[A_{ij},B_{ji}] )
\nn\\
 &=&\mbox{ / using $[A,B]=0$ / }
= \sum_j k_j (\sum_i
[B_{ij},A_{ji}]-[A_{ij},B_{ji}] )
\nn\\
&=& \sum_j k_j
(\sum_i [B_{ij},A_{ji}]-[A_{ij},B_{ji}] ) = \sum_j k_j (\sum_i [ \one
B , \two A]_{ij,ji} -[ \one A, \two B]_{ij,ji}  ) =
\nn\\&=&
 \sum_j \sum_i k_j ( ([
\one B , \two A]P)_{ii,jj} -([ \one A, \two B]P)_{ii,jj}  ) = Tr \two K (
([ \one B , \two A]P) -([ \one A, \two B]P)  ) = \nn\\&=&
Tr \two K ( ([ \one B ,
\two A]P) -(P[ \two A, \one B])  ) =
Tr \two K ( ([ \one B , \two A]P)
+(P[ \one B , \two A ])  )
\nn\\&=&
Tr P (\two K + \one K ) ([ \one B , \two A] ) \nn
.\eea
$\Box$

{\Lem ~ \label{TrKPhi-TrK2Phi}
\bea
[ Tr
L_{MF}^{2}(u),    Tr  \K^2    L_{MF}^{n}(z)  ]=
\frac{2}{u}[ Tr \K \Phi,  Tr  \K^2    L_{MF}^{n}(z)  ]=
0
\eea
}
\\
{\bf Proof~} Let us note that $$ Tr L_{MF}^{2}(u) = Tr \K^2 + \frac{2}{u} Tr \K \Phi
+\frac{1}{u^2}Tr \Phi^2.$$ Recalling that $Tr \K^2, Tr \Phi^2$ commute with
everything  we see that it is enough to prove
$$[ Tr \K\Phi ,    Tr  \K^2    L_{MF}^{n}(z)  ]=0$$
or equivalently $$
[ Tr \K L_{MF}(u) ,    Tr  \K^2    L_{MF}^{n}(z)  ]=0.$$
So let us calculate:
\bea&&
[ Tr  \K^2    L_{MF}^{n}(z), Tr \K L_{MF}(u) ]=
Tr [ \one \K^2 \one L_{MF}^{n}(z)), \two K \two L_{MF}(u) ]
\nn\\
&=&
Tr \one \K^2 \two \K   [ \one L_{MF}^{n}(z), \two L_{MF}(u) ]
=
\mbox{ / by lemma \ref{ln-l-easy} and formula \ref{fml-ln-l-easy} /}
\nn\\
&=&
Tr \one \K^2 \two \K
(\two L^n(z)  - \one L^n(z) ) \frac{P}{z-u}+ Tr \one \K^2 \two \K
\sum_{i=0}^{n-1} \one L^i(z) (\one L(u) - \two L(u)) \two L^{n-1-i}(z)
\frac{P}{z-u}
\nn \\
&=&
Tr \one \K^2 \two \K
\sum_{i=0}^{n-1} \one L^i(z) (\frac{z}{u}\one L(z) -\one \K(\frac{z}{u}-1) -
\frac{z}{u}\two L(z) +\two \K (\frac{z}{u}-1)) \two L^{n-1-i}(z)
\frac{P}{z-u}
\nn
\eea
\bea
&=&
Tr \one \K^2 \two \K
\sum_{i=0}^{n-1} \one L^i(z) (\frac{z}{u}\one L(z)  -
\frac{z}{u}\two L(z) ) \two L^{n-1-i}(z)
\frac{P}{z-u}
\nn \\
&=&
\frac{z}{u} Tr \one \K^2 \two \K (
\one L^{n}(z)
\frac{P}{z-u}-
\two L^{n}(z)
\frac{P}{z-u}
)
\nn \\
&=&
\frac{z}{u} (Tr \K^2  L^{n}(z) \K
-
 Tr \K^2 \K L^{n}(z))
\frac{1}{z-u}
)=0 \nn
\eea
$\Box$

{\Lem
\label{faf-bf}
 $$[Tr \Phi A \Phi, Tr B \Phi ]=0, \mbox{~~for~~} A,B: [A,B]=0$$}
\\
{\bf Proof~} By a straightforward calculation.

{\Lem \label{phin-k-phi}
$$ Tr \Phi^n \K \Phi - \Phi \K \Phi^n = 0. $$
}
\\
{\bf Proof~}
According to the formula \ref{BKA-AKB}:
\bea
&&Tr \Phi^n \K \Phi - \Phi \K \Phi^n =
Tr P (\two \K + \one \K )
([ \one \Phi^n , \two  \Phi ] )
=
Tr P (\two \K + \one \K )
([ \one \Phi^n , P ] )
\nn\\
&=&
Tr P (\two \K + \one \K )
( \one \Phi^n  P -  \two \Phi^n  P  )
=
Tr  (\two \K   \one \Phi^n   - \two \K   \two \Phi^n    )
+( \one \K  \one \Phi^n   -  \one \K \two \Phi^n    )
\nn\\
&=& Tr  \K   Tr  \Phi^n   -  Tr Id Tr \K   \Phi^n
+  Tr \K \Phi^n Tr Id    - Tr \K  Tr\Phi^n    )
=0\nn
\eea
$\Box$

{\Lem ~~ \label{lem-phi-k-phi}
\bea
Tr [\Phi, \K^l \Phi^m]=  Tr  \K^l  ~~  Tr  \Phi^m-  Tr  \K^l  \Phi^m ~~ Tr Id
\eea
}
\\
{\bf Proof}
\bea
Tr [\Phi, \K^l \Phi^m]&=&
  Tr [\one \Phi, \two \K^l \two \Phi^m] P
=
 Tr \two \K^l [\one \Phi,  \two \Phi^m] P
=  Tr \two \K^l [ P,  \two \Phi^m ] P
=
\nn\\
&=&
  Tr \two \K^l \one \Phi^m-  Tr \two \K^l \two \Phi^m
=
 Tr  \K^l  ~~  Tr  \Phi^m-  Tr  \K^l  \Phi^m ~~ Tr Id
\eea
$\Box$

%
%

{\Lem ~~ \label{l-Skr}}  (due to
Skrypnyk \cite{Skr}):
\bea
\label{fml-Skr}
[  Tr  A  \Phi ,
  Tr B \Phi^n ] = 0 \mbox{~~ where~~ } [A,B]=0
\eea
\\
{\bf Proof}:
\bea
Tr [\one A \one \Phi, \two B \two \Phi^n ]
&=&
Tr \one A  \two B [ P ,  \two \Phi^n ]
\nn\\&=&
Tr \one A  \two B    \one \Phi^n P
-
Tr \one A  \two B   \two \Phi^n P
= Tr  A      \Phi^n B - Tr  A B      \Phi^n =
0
\eea
$\Box$

{\Lem ~~ \label{lem-k-phi-n}}
\bea
\label{k-phi-n}
[Tr (2\K^2\Phi^2+\K\Phi\K\Phi),
   Tr  \K  \Phi^n  ] = 0
\eea
\\
{\bf Proof~}
\bea
&&[Tr \K^2\Phi^2,
  Tr \K \Phi^n]=
Tr [\one  \K^2 \one\Phi^2,   \two  \K \two \Phi^n]
= Tr \one  \K^2 \two  \K [\one \Phi^2,   \two \Phi^{n}]
\nn\\ &=&
\mbox{/ by lemma \ref{l3} /}
=Tr \one  \K^2 \two  \K (\sum_{a=1,2} P \one \Phi^{a-1}  \two \Phi^{n+2-a}
- P \one \Phi^{n+2-a}  \two \Phi^{a-1} )
\nn\\
&=&Tr \one  \K^2 \two  \K (P   \two \Phi^{n+1}
- P \one \Phi^{n+1}   +
P \one \Phi  \two \Phi^{n}
- P \one \Phi^{n}  \two \Phi)
=Tr \one  \K^2 \two  \K P (
 \one \Phi  \two \Phi^{n}
-  \one \Phi^{n}  \two \Phi)
\nn\\
&=& Tr \two  \K  \two \Phi   \one  \K^2  \one \Phi^{n} P
- Tr \two  \K  \two  \Phi^{n}   \one  \K^2  \one \Phi P
=Tr  \K
   \Phi     \K^2   \Phi^{n}
- Tr  \K  \Phi^{n}   \K^2  \Phi
\eea

\bea
&&[ Tr \K\Phi\K\Phi,
  Tr \K\Phi^n ] =
Tr \one \K \one \Phi \one \K \two \K [  \one \Phi,
\two \Phi^n] + \one \K \two \K [  \one \Phi,
\two  \Phi^n] \one \K \one \Phi
\nn\\&=&
\mbox{/ by lemma \ref{l2} /}  \nn\\
&=&
Tr \one \K \one \Phi \one \K \two \K \one \Phi^n P
- \one \K \one \Phi \one \K \two \K \two \Phi^n P
+
\one \K \two \K   \one \Phi^n \two \K \two \Phi P
\one \K \two \K  \two \Phi^n \two \K \two \Phi P
\nn\\
&=&
Tr \K^2  \Phi \K  \Phi^n
- \K \Phi \K^2\Phi^n
+
\K \Phi^n \K^2 \Phi -
\K^2\Phi^n\K \Phi
\nn\\
&=&
Tr [\K^2  \Phi, \K  \Phi^n]
-2 \K \Phi \K^2\Phi^3
+
2 \K \Phi^n \K^2 \Phi -
[\K^2\Phi^n, \K \Phi ]
\nn\\
&=&
2 Tr  \K \Phi^n \K^2 \Phi
- 2 Tr \K \Phi \K^2\Phi^n
+ Tr [\K^2  \Phi, \K  \Phi^n]
- Tr[\K^2\Phi^n, \K \Phi ]
%
\eea

\bea
Tr [\K^2  \Phi, \K  \Phi^n]&=&
Tr \one \K^2 \two \K  [P, \two \Phi ^n] P
=
Tr \one \K^2 \two \K   \one \Phi ^n -
Tr \one \K^2 \two \K   \two \Phi ^n
\nn\\ &=&
Tr \K^2  \Phi ^n Tr \K -
Tr  \K^2 Tr  \K \Phi ^n
\\
Tr [\K^2  \Phi^n, \K  \Phi]&=&
Tr \one \K^2 \two \K  [\one \Phi ^n, P] P=
Tr \one \K^2 \two \K   \one \Phi ^n -
Tr \one \K^2 \two \K   \two \Phi ^n
\nn\\
&=&
Tr \K^2  \Phi ^n Tr \K -
Tr  \K^2 Tr  \K \Phi ^n
\eea

Hence
\bea
Tr [\K^2  \Phi, \K  \Phi^n]=
Tr [\K^2  \Phi^n, \K  \Phi]
\eea

\bea
&&[ Tr \K\Phi\K\Phi,
  Tr \K\Phi^n ] =
2 Tr  \K \Phi^n \K^2 \Phi
- 2 Tr \K \Phi \K^2\Phi^n
%
\eea

\bea
[Tr (2\K^2\Phi^2+\K\Phi\K\Phi),
   Tr  \K  \Phi^n  ]=0
\eea
The lemma is proved. $\Box$


\end{document}